\documentclass[11pt]{article}
\usepackage{lmodern}
\usepackage[T1]{fontenc}
\usepackage[utf8]{inputenc}
\usepackage[margin=1in]{geometry}
\usepackage{amsmath,amssymb,amsthm}
\usepackage{graphicx}
\usepackage{tabularx}
\usepackage[hidelinks]{hyperref}
\usepackage{microtype}
\usepackage[font=small,labelformat=empty]{caption}
\title{An Honest Effect Size for Contingency Tables:\\ Why Nothing Can Be Unbiased, Where to Put the Error Instead, and How to Route the Report}
\author{William J. Dwyer, MD, MPH, FAAP\\ Department of Mathematics and Statistics,\\ University of Massachusetts Lowell, Lowell, MA, USA\\ \texttt{wjdwyer@trialdesign.com} \\ ORCID 0009-0004-0855-7222}
\date{}
\begin{document}
\maketitle

\begin{abstract}
\textbf{Background.} Cramer's V is the effect size reported beside almost every chi-square test, as a bare number read against Cohen's labels, yet a large fraction of such numbers describe sampling noise and the standard bias correction (Bergsma 2013) does not fix the problem. We ask whether it works, and what would. \textbf{Methods.} We assemble three results and claim only the consequence of each. First, the squared effect size $\phi$\textsuperscript{2} admits no unbiased estimator at any sample size, because under fixed-N multinomial sampling the expectation of any estimator is a polynomial in the cell probabilities while $\phi$\textsuperscript{2} is not. Second, unbiasedness transfers across a rescaling of the effect size if and only if the rescaling is affine, and among the affine choices V\textsuperscript{2} = $\phi$\textsuperscript{2}/k is the one bounded in [0,1] with value 1 at perfect association. Third, we report an interval with conservative, asymptotically valid coverage obtained by projecting a likelihood-ratio confidence set for the cell probabilities through the effect-size map; its lower endpoint is zero in closed form exactly when the test of independence fails to reject. All estimators are evaluated against known truth by simulation with Monte Carlo standard errors. \textbf{Results.} Because nothing is unbiased, the only question is where the irreducible error is placed. Bergsma's correction puts zero error at the null and several percent under the alternative; a delete-one jackknife on the V\textsuperscript{2} scale spreads it thin everywhere (absolute bias at most 0.008 across the 180-design grid). Pooling does not remove bias: across 3,000 simulated meta-analyses, pooling 200 studies drives the naive estimator's probability of landing within 0.01 of the truth to zero while the jackknife's rises to 0.99. The projected interval covers 0.997 to 1.000 across the tested designs while the noncentral inversion undercovers (0.936 at $\phi$\textsuperscript{2} = 0.18), at two to three times the width. \textbf{Conclusions.} The point estimate and the interval are different problems with different answers, and conflating them is why the literature has neither. We give a routing rule and a browser tool that implements it: the reference follows sparsity, the verdict follows the noise floor, the point estimate follows what the number is for, the scale is always V\textsuperscript{2}, and the interval follows whether a guarantee is required.
\end{abstract}


\section{Introduction}

A researcher cross-tabulates two categorical variables, runs a chi-square test, and reports Cramer's V against Cohen's labels. A companion paper (Dwyer, 2026d) shows how often that number is indistinguishable from chance: across 4,129 real two-way tables, 21.8 percent of the reported Cohen labels sit on tables with no significant association at all.

The obvious remedy is the bias correction of Bergsma (2013), which is in every major package. \textbf{This paper is about what happens when you try to use it, and what to do instead.}

\subsection{What we found when we tried to fix it}

We set out to build a better bias correction, and the attempt failed in an instructive way. Three things emerged, in this order.

\textbf{First, the correction we tried to improve on is already known to be imperfect, and Bergsma says so himself.} His Section 3 exists precisely to ask whether the correction survives dependence, and his Figures 1 through 3 plot the residual bias against the strength of association. We claim none of this. What we contribute is the reason and the remedy.

\textbf{Second, and this is the organizing fact of the paper: there is no unbiased estimator to build.} Not a hard one. Not an expensive one. \textbf{None, at any sample size} (Section 3). This is not a limitation of our effort; it is a theorem, and it is elementary once stated.

\textbf{Third, once that is accepted, the entire question changes.} If the error cannot be removed, it can only be \textbf{placed}, and where to place it depends on what the number is going to be used for. That is a routing problem, and it is the shape of the answer.

\subsection{The classical apparatus, stated fairly}

With row sums r\_i, column sums c\_j, total N, expectations E\_ij = r\_i c\_j / N, and

\[
\phi ^{2} = X^{2} / N, ,\quad dof = (R-1)(C-1), ,\quad k = \min (R-1, C-1),
\]

the family is

\[
Cohen's w = \surd (\phi ^{2}) \\
Cramer's V = \surd (\phi ^{2} / k) \\
Tschuprow's T = \surd (\phi ^{2}/\surd (dof)) \\
Pearson's C = \surd (\phi ^{2}/(1+\phi ^{2})).
\]

\textbf{Every one of them is a square root of a rescaling of $\phi$\textsuperscript{2}.} The field took a root five separate ways and never asked whether it needed one. That question turns out to matter a great deal (Section 4).

The null expectation

\[
E[\hat{\phi }^{2}] = dof / (N - 1) ,\quad under independence, multinomial sampling,
\]

was conjectured by Tschuprow (1925) and \textbf{proven by Bartlett (1937)}. Bergsma (2013) supplies the correction $\hat{\phi}$\textsuperscript{2} - dof/(N-1), truncated at zero, together with adjusted denominators r\textasciitilde{} and c\textasciitilde{} so that a perfect association still maps to 1. \textbf{All of this is correct, and all of it is theirs.}

\subsection{What is new here, and what is not}

This paper claims no new mathematics, and it is better to say so before saying anything else. An adversarial prior-art audit, deposited with the code, refuted three of the four contributions an earlier draft claimed, and the table below records the outcome rather than burying it in a footnote.

\begin{table}[htbp]\centering\small
\begin{tabularx}{\textwidth}{>{\raggedright\arraybackslash}X>{\raggedright\arraybackslash}X>{\raggedright\arraybackslash}X}
\hline
Strand & Whose it is & What we add \\
\hline
E[$\phi$\textsuperscript{2}] = dof/(N-1) at independence & Tschuprow (1925); Bartlett (1937) & Nothing \\
The bias correction & Bergsma (2013) & Nothing. It is our comparator. \\
Residual bias of that correction under the alternative & Bergsma (2013), Section 3, Figures 1 to 3 & Nothing. \textbf{It is his own finding and we claim none of it.} \\
No unbiased estimator exists & Girshick, Mosteller and Savage (1946); Lehmann and Casella, Section 2.1; stated verbatim for entropy and mutual information by Paninski (2003, Proposition 8) & The one-line consequence for $\phi$\textsuperscript{2}, which this literature appears never to have recorded. A remark, not a theorem. \\
The delete-one jackknife & Quenouille (1949); Tukey (1958); Fay (1985), who jackknifed Pearson's X\textsuperscript{2} & The instantiation on the V\textsuperscript{2} scale, a closed form, and a measurement; Fay's X\textsuperscript{2} jackknife is the closely related precedent \\
The projected interval & \textbf{Scheffe (1953); Stark's strict bounds; Kaido, Molinari and Stoye (2019); confidence-procedure marginalization} & \textbf{The instantiation for $\phi$\textsuperscript{2} and Cramer's V, a lower endpoint that is zero in closed form exactly at non-rejection, and the benchmark showing that every interval fails somewhere across the designs benchmarked here and the 180-cell grid} \\
Lower endpoint is zero exactly when the test fails to reject & Test and interval duality & Nothing beyond the instantiation. An earlier draft called this a finding. It is not. \\
\hline
\end{tabularx}
\end{table}

What is left is a measurement and an assembly, and we think that is enough. Specifically: unbiasedness on $\phi$\textsuperscript{2} transfers only through an affine link, and V\textsuperscript{2} = $\phi$\textsuperscript{2}/k is the affine member bounded in [0,1] (Section 4); \textbf{bias does not average away}, and pooling 200 studies drives the naive estimator's hit rate to zero, which appears never to have been quantified for this family (Section 5); the square root breaks the point estimate and leaves the interval untouched, so they are different problems with different answers (Section 6); and the five decisions are routed and made consistent by construction (Section 7).

\section{The classical approach and why it disappoints}

The received practice is: compute X\textsuperscript{2}, compute V, look up Cohen. If the analyst is careful, apply Bergsma's correction. \textbf{Figure 1 shows what that produces when the true effect is V = 0.10}, across twelve shapes and five sample sizes.

\begin{figure}[htbp]\centering
\includegraphics[width=\linewidth]{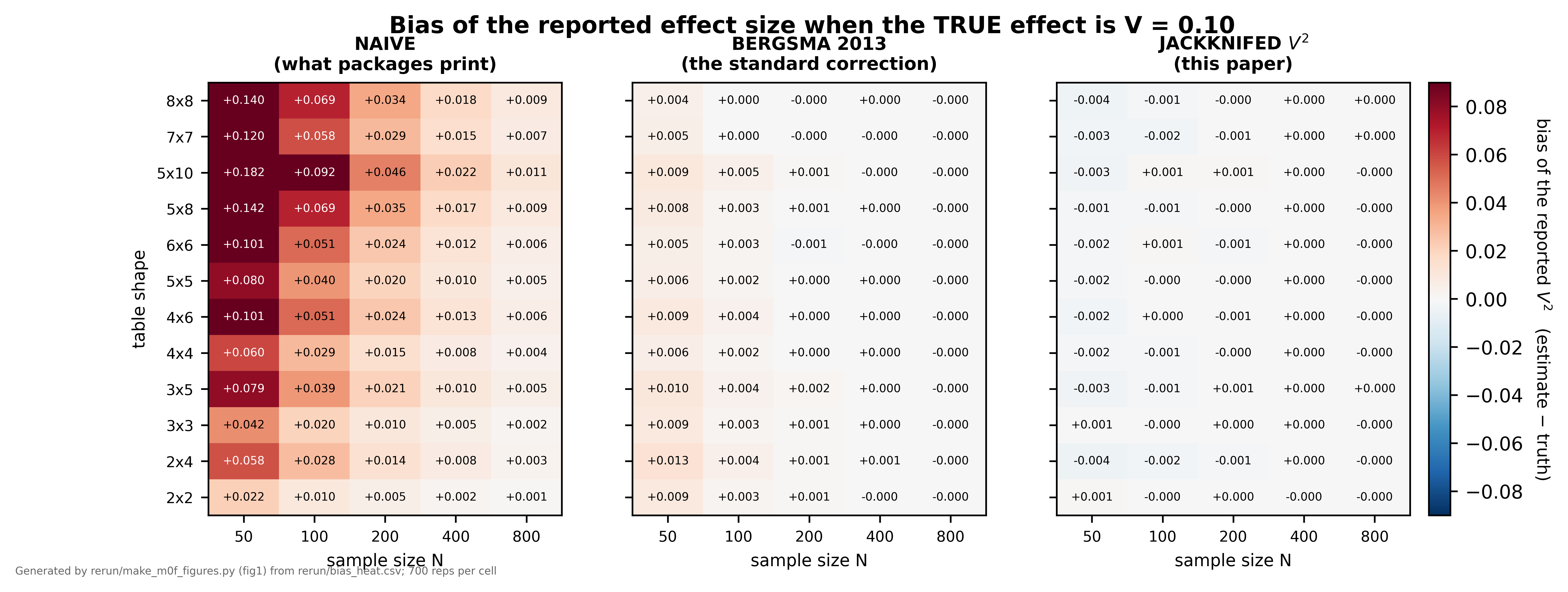}
\caption{\textbf{Figure 1.} Bias on the V\textsuperscript{2} scale of three $\phi$\textsuperscript{2} estimators at a true effect of V = 0.10, across twelve table shapes and five sample sizes. Left: the naive plug-in, whose bias reaches +0.18 in the sparsest corner. Center: Bergsma's correction, whose residual here is small (no more than about 0.013). Right: the delete-one jackknife on the V\textsuperscript{2} scale, within 0.008 everywhere. Generated by rerun/make\_m0f\_figures.py.}
\end{figure}

The naive estimator's bias reaches \textbf{+0.182 on the V\textsuperscript{2} scale} in the sparsest corner, which is to say the reported effect is many times the real one. Bergsma's correction removes most of that, and this is a real and useful achievement that we do not dispute. But its residual bias here is small, no more than about \textbf{0.013} on the V\textsuperscript{2} scale; what matters is not its size in this slice but where the correction places its error, since Bergsma's is built to vanish at the null and leaves its residual under the alternative, which is where something is at stake (Section 5). The jackknifed V\textsuperscript{2} stays within \textbf{0.008 everywhere}.

The natural response is: so build a better correction. We tried. Section 3 explains why that road ends.

\section{Nothing is unbiased. This is standard, and it is the organizing fact.}

\textbf{Attribution first.} This section contains no new mathematics and we claim none. The argument is the classical U-estimability criterion of Girshick, Mosteller and Savage (1946), set out in Lehmann and Casella (Theory of Point Estimation, Section 2.1), and Paninski (2003, Proposition 8) states it verbatim for entropy and mutual information under exactly this fixed-N multinomial model. Mutual information is the log-divergence sibling of $\phi$\textsuperscript{2}. What appears not to have been recorded is the one-line consequence for $\phi$\textsuperscript{2}, and hence for Cramer's V, which the bias-correction literature from Tschuprow (1925) to Bergsma (2013) never states. We record it as a remark. Nothing in this paper depends on its being new.

Let O \textasciitilde{} Multinomial(N, p) on the R $\times$ C table. For \textbf{any} estimator T,

\[
E[T(O)] ,\quad = ,\quad \sum over tables O ,\quad \text{of} ,\quad T(O) * N!/(prod o_{ij}!) * prod p_{ij}^{o_{ij}}.
\]

The right-hand side is a \textbf{polynomial in p of degree exactly N}. This holds for every T, without exception, because it is simply the multinomial expectation. A function of p is therefore unbiasedly estimable only if it is a polynomial on the simplex of degree at most N (Girshick, Mosteller and Savage 1946; Lehmann and Casella, Section 2.1).

\textbf{Is $\phi$\textsuperscript{2} a polynomial? No, and the proof is three lines in the interior of the simplex.} Fix c in (0, 1/2) and take the curve p(t) = [[t, c], [c, 1 - t - 2c]] for t in (0, 1 - 2c). Both margins are (t + c, 1 - t - c), so with D(t) = t(1 - t - 2c) - c\textsuperscript{2},

\[
\phi ^{2}(t) ,\quad = ,\quad D(t)^{2} / [ (t + c)(1 - t - c) ]^{2}.
\]

A polynomial is entire. This function has a pole at t = -c, because (t + c) does not divide D(t): indeed D(-c) = -c, which is not zero. Therefore $\phi$\textsuperscript{2} is not a polynomial, and:

\textbf{Under fixed-N multinomial sampling, no unbiased estimator of $\phi$\textsuperscript{2} exists, for any N.}

The qualifier is not optional. Under inverse or sequential sampling plans, non-polynomial functions do become unbiasedly estimable, which is itself part of what Girshick, Mosteller and Savage (1946) establish.

\textbf{A correction we owe the reader.} An earlier draft of this work called the statement above a theorem, attributed the polynomial characterization to Halmos (1946), and proved non-polynomiality by exhibiting different limits at a \textbf{vertex} of the simplex. All three were wrong. Halmos treats unbiased estimation of regular functionals in the nonparametric i.i.d. model and does not contain the fixed-N multinomial identity. And $\phi$\textsuperscript{2} is \textbf{not defined} at those vertices, since a zero margin gives 0/0, so on the open simplex, where $\phi$\textsuperscript{2} actually lives, that proof established nothing. The interior argument above replaces it, and the correct citations replace Halmos.

Bartlett's exact result is not a counterexample: it gives the value of E[$\hat{\phi}$\textsuperscript{2}] \textbf{at a point} (independence), not unbiasedness as a function of p.

\subsection{What the theorem does to the problem}

\textbf{It replaces the question.} "Is the estimator unbiased?" has only one answer, and it is no. The real question is:

\textbf{Given that a bias must exist somewhere, WHERE should it be put?}

That is a design decision, and it has a right answer only relative to a purpose. \textbf{This is the argument for routing, and it is a theorem-level argument, not a matter of taste.}

\section{The scale: only an affine link survives}

Suppose an estimator J is (nearly) unbiased for $\phi$\textsuperscript{2}, and we want to report g($\phi$\textsuperscript{2}) instead. When does unbiasedness transfer?

\textbf{Expectation commutes with affine maps and with nothing else.} E[g(J)] = g(E[J]) for all distributions \textbf{if and only if g is affine.} Every nonlinear g introduces a Jensen term that no bias correction on the $\phi$\textsuperscript{2} scale can anticipate.

We tested this rather than asserting it. Eight links, jackknifed on each link's own scale, 5$\times$5 at N = 100, 1,500 reps, bias measured against g($\phi$\textsuperscript{2}\_true) at the null.

\begin{figure}[htbp]\centering
\includegraphics[width=\linewidth]{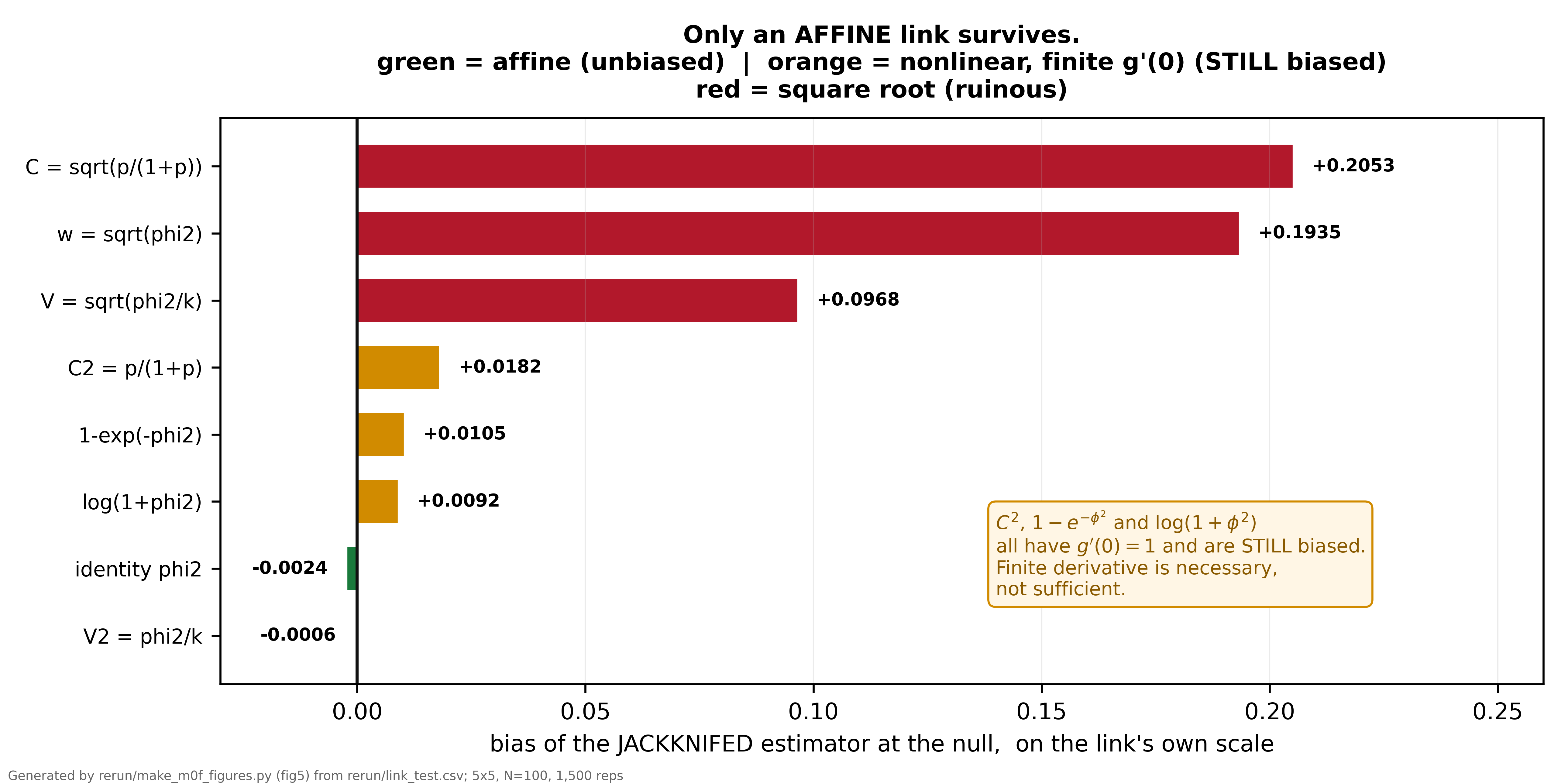}
\caption{\textbf{Figure 2.} Bias of eight effect-size links, each jackknifed on its own scale, at the null (5$\times$5, N = 100, 1,500 replications). Only the affine links ($\phi$\textsuperscript{2} and V\textsuperscript{2} = $\phi$\textsuperscript{2}/k) are unbiased; the square-root links (V, w, C) and the other nonlinear links carry the largest bias. Generated by rerun/make\_m0f\_figures.py.}
\end{figure}

\begin{table}[htbp]\centering\small
\begin{tabular}{lcc}
\hline
link & g'(0) & bias at the null \\
\hline
\textbf{identity, $\phi$\textsuperscript{2}} & 1 & \textbf{-0.0024 $\pm$ 0.0016} \\
\textbf{V\textsuperscript{2} = $\phi$\textsuperscript{2} / k} & 1 & \textbf{-0.0006 $\pm$ 0.0004} \\
C\textsuperscript{2} = $\phi$\textsuperscript{2}/(1+$\phi$\textsuperscript{2}) & 1 & +0.0182 $\pm$ 0.0015 \\
1 - exp(-$\phi$\textsuperscript{2}) & 1 & +0.0105 $\pm$ 0.0016 \\
log(1 + $\phi$\textsuperscript{2}) & 1 & +0.0092 $\pm$ 0.0015 \\
\textbf{V = $\surd$($\phi$\textsuperscript{2}/k)} & \textbf{infinite} & \textbf{+0.0968 $\pm$ 0.0015} \\
\textbf{w = $\surd$($\phi$\textsuperscript{2})} & \textbf{infinite} & \textbf{+0.1935 $\pm$ 0.0029} \\
C = $\surd$($\phi$\textsuperscript{2}/(1+$\phi$\textsuperscript{2})) & infinite & +0.2053 $\pm$ 0.0026 \\
\hline
\end{tabular}
\end{table}

\textbf{We had expected the criterion to be a finite derivative at zero. It is not.} C\textsuperscript{2}, 1 - exp(-$\phi$\textsuperscript{2}) and log(1+$\phi$\textsuperscript{2}) all have g'(0) = 1 and all three remain biased. \textbf{Finite derivative is necessary and not sufficient. Only affine works}, and

\textbf{V\textsuperscript{2} = $\phi$\textsuperscript{2} / k is the affine member of the chi-square family bounded in [0, 1].}

$\phi$\textsuperscript{2} itself is trivially affine, being the identity, so affinity alone does not single out any scale; what distinguishes V\textsuperscript{2} among the affine rescalings is the [0, 1] normalization. It is bounded in [0, 1]. It equals 1 at perfect association for any R $\times$ C. It is linear in $\phi$\textsuperscript{2}, so unbiasedness transfers exactly. \textbf{And it may be negative, which is not a defect: it is precisely what adjusted R\textsuperscript{2} does, and it carries the same meaning: the observed association is weaker than chance alone would have produced.}

\subsection{And the square root is not always fatal}

At $\phi$\textsuperscript{2} = 0.64 the jackknifed V is biased by \textbf{-0.0001, which is -0.02 percent.} The jackknife removes the O(1/N) bias of any smooth functional, and the square root is perfectly smooth away from zero.

\textbf{The square root is ruinous in a neighborhood of the null and, in the case tested, harmless away from it.} V is safe for a strong association and misleading for a weak one, \textbf{and the researcher who most needs to be told the truth is the one holding a weak one.}

\section{The point estimate: where to put the error}

\begin{figure}[htbp]\centering
\includegraphics[width=\linewidth]{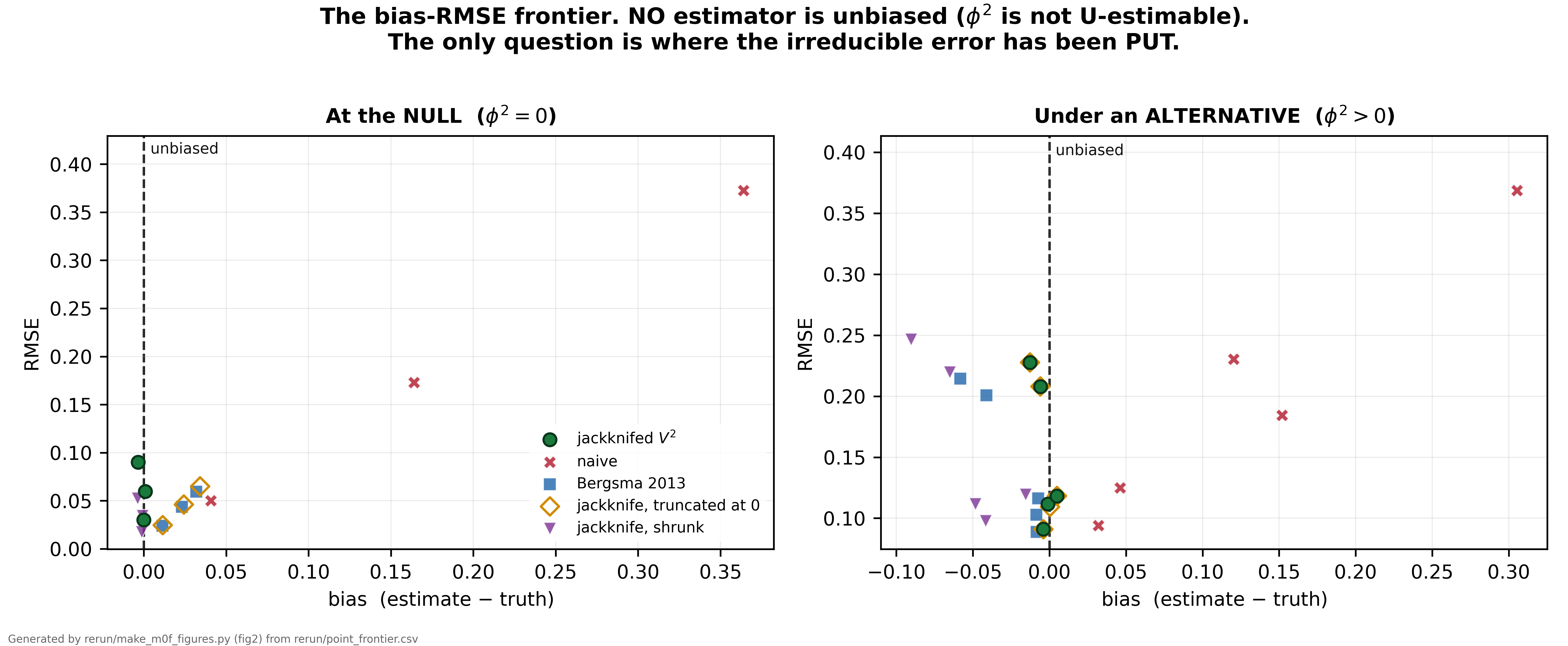}
\caption{\textbf{Figure 3.} The bias-RMSE frontier for $\phi$\textsuperscript{2} estimators (5$\times$5, N = 100): naive, Bergsma, and the delete-one, truncated, and shrunk jackknives, plotted by absolute bias against root mean squared error at the null and under a moderate alternative. No estimator is unbiased; the frontier shows where each places its irreducible error. Generated by rerun/make\_m0f\_figures.py.}
\end{figure}

Since nothing is unbiased, the estimators lie on a frontier. Five candidates, measured against known truth:

\begin{table}[htbp]\centering\small
\begin{tabular}{lccc}
\hline
estimator & bias at null & bias under alternative & RMSE at null \\
\hline
naive & +0.164 & +0.120 & 0.173 \\
Bergsma 2013 & +0.023 & \textbf{-0.041} & \textbf{0.044} \\
\textbf{jackknifed V\textsuperscript{2}} & \textbf{+0.001} & \textbf{-0.006} & 0.060 \\
jackknife, truncated at 0 & +0.024 & -0.006 & 0.046 \\
jackknife, shrunk & -0.001 & -0.065 & \textbf{0.034} \\
\hline
\end{tabular}
\end{table}

(5$\times$5, N = 100; alternative at $\phi$\textsuperscript{2} = 0.64. Full grid in \texttt{point\_frontier.csv}.)

The jackknife applied to a chi-square statistic is not itself new. Fay (1985) jackknifed Pearson's X\textsuperscript{2}, and Jiao et al. (2017) give jackknife bias correction for plug-in functionals of a discrete distribution, of which $\phi$\textsuperscript{2} is a special case; and rcompanion and DescTools already form delete-one jackknife pseudo-values of Cramer's V inside their BCa bootstrap. The pseudo-values exist in that software; what is new here is the bias-correction framing on the V\textsuperscript{2} scale, the closed form, and the measurement. A separate line of work, the Bayesian bias-corrected estimators of Momozaki et al. (2024), is a direct competitor, reaching the same goal of a lower-bias estimator of the generalized Cramer coefficient by a different route.

\textbf{Three readings, and the third is the most useful.}

\textbf{The truncation is the culprit, not the correction.} "jackknife, truncated at 0" is the jackknife with max(0, .) applied, and it recovers \textbf{exactly the null bias that Bergsma has} (+0.024 against +0.023). The non-negativity constraint, not the choice of correction, is what manufactures the bias at the null. \textbf{This is the strongest single argument for reporting V\textsuperscript{2} and permitting negatives.}

\textbf{Shrinkage buys RMSE and sells alternative-bias.} The shrunk jackknife has the best RMSE at the null of anything we tested (0.034, better than Bergsma's 0.044), and the worst bias under the alternative (-0.065). It is a legitimate point on the frontier, and we report it as such rather than as a free lunch.

\textbf{No estimator dominates.} That is the theorem made visible.

\subsection{The demonstration that decides it: bias does not average away}

An effect size is rarely used alone. It is pooled, and it powers the next study. \textbf{Both operations average. Does the error wash out?}

\begin{figure}[htbp]\centering
\includegraphics[width=\linewidth]{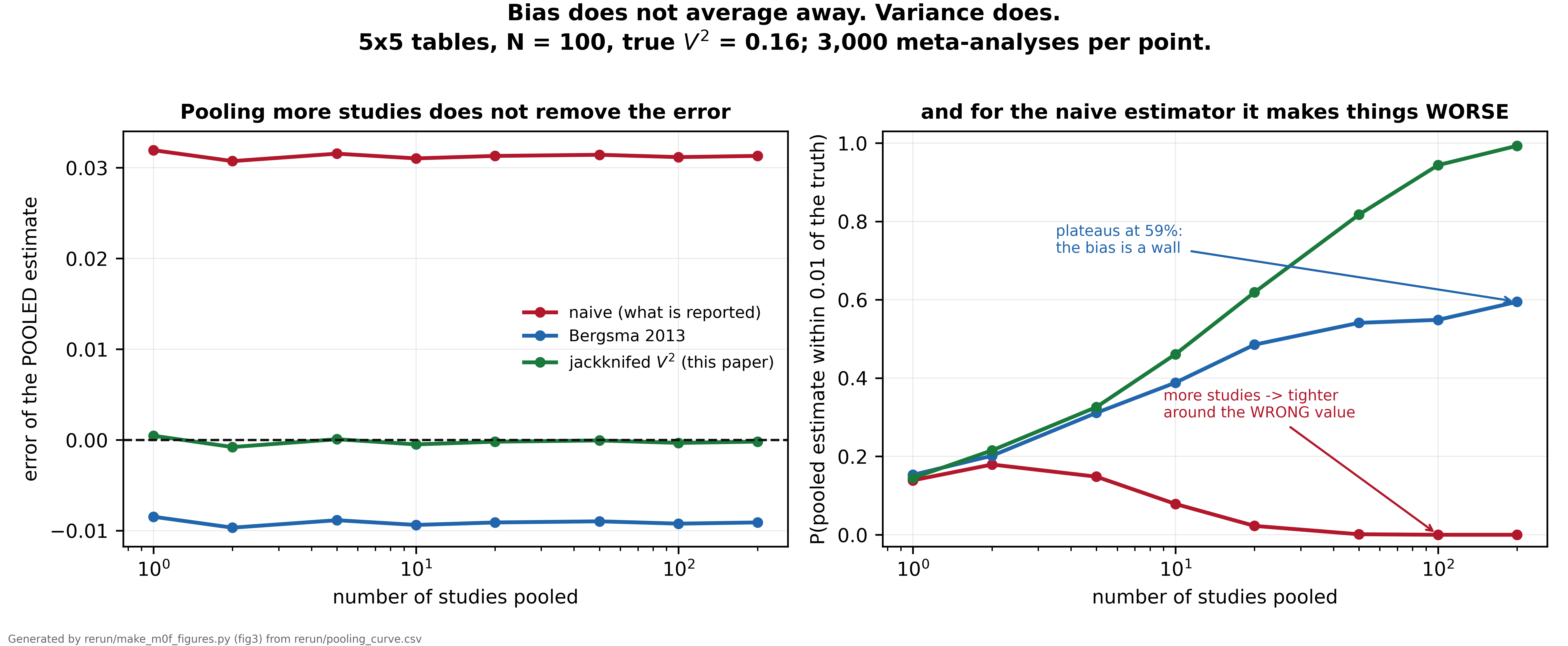}
\caption{\textbf{Figure 4.} Pooling does not remove bias. Probability that a pooled estimate lands within 0.01 of the truth as the number of pooled studies grows (3,000 simulated meta-analyses). The naive estimator's hit rate falls to zero, Bergsma's plateaus near 0.57, and the delete-one jackknife's rises toward 0.99. Generated by rerun/make\_m0f\_figures.py.}
\end{figure}

3,000 simulated meta-analyses, 5$\times$5 tables at N = 100, true V\textsuperscript{2} = 0.16:

\begin{table}[htbp]\centering\small
\begin{tabular}{lccc}
\hline
studies pooled & error: naive & error: Bergsma & error: \textbf{jackknife} \\
\hline
1 & +0.0324 & -0.0080 & \textbf{+0.0009} \\
10 & +0.0308 & -0.0096 & \textbf{-0.0007} \\
50 & +0.0311 & -0.0093 & \textbf{-0.0004} \\
200 & +0.0310 & -0.0094 & \textbf{-0.0005} \\
\hline
\end{tabular}
\end{table}

\textbf{Nothing improves.} And the probability that the pooled estimate lands within 0.01 of the truth:

\begin{table}[htbp]\centering\small
\begin{tabular}{lccc}
\hline
studies pooled & naive & Bergsma & \textbf{jackknife} \\
\hline
1 & 0.146 & 0.159 & 0.147 \\
10 & 0.085 & 0.388 & \textbf{0.456} \\
50 & 0.002 & 0.529 & \textbf{0.827} \\
200 & \textbf{0.000} & 0.568 & \textbf{0.995} \\
\hline
\end{tabular}
\end{table}

\textbf{Read the naive column downward. Pooling more studies makes it WORSE}, from 0.146 to 0.000, because the variance shrinks around the wrong center. \textbf{A meta-analysis of the reported literature converges, with confidence, on a value that is not there.}

\textbf{Bergsma's plateaus at 0.57.} More data does not help, because the residual is a bias and biases are walls.

\textbf{The jackknife converges to 0.995.}

\textbf{This is why the point estimate must be routed by purpose.} When the number will be averaged, the loss is bias. When it describes one table and will not be averaged, the loss is squared error, and Bergsma minimizes that. \textbf{Neither is wrong. Using the wrong one is.}

\subsection{The verdict flips both ways, on real published tables}

The frontier and the pooling demonstration are simulations. The distortion is visible directly in the published record, and it runs in \textbf{both directions}, because a bare V read against a Cohen benchmark ignores the one quantity the noise floor accounts for: the sample size. A scan of 4,129 real two-way tables from public datasets (the companion noise-floor paper, Dwyer 2026d) finds \textbf{630 whose Cohen V-label changes once the bias is corrected, and 1,711 whose reported effect does not clear its own noise floor.} The ten tables below, each from a distinct public dataset, show the two failure directions; every value is reproduced by the deposited engine.

\textbf{When the table is small, the benchmark is too generous.} The naive V earns a "medium" or "large" Cohen label, but the bias-corrected and jackknife estimates fall toward zero, the value sits below its noise floor, and the $\chi$\textsuperscript{2} test does not reject: the magnitude is sampling noise wearing a label.

\begin{table}[htbp]\centering\small
\begin{tabularx}{\textwidth}{>{\raggedright\arraybackslash}X>{\raggedright\arraybackslash}X>{\raggedright\arraybackslash}X>{\raggedright\arraybackslash}X>{\raggedright\arraybackslash}X>{\raggedright\arraybackslash}X>{\raggedright\arraybackslash}X>{\raggedright\arraybackslash}X}
\hline
Table (field) & shape & N & $\chi$\textsuperscript{2} p & naive V (Cohen) & Bergsma & jackknife & floor V\_0.95 \\
\hline
Mammal diet $\times$ conservation status (ecology) & 4$\times$6 & 52 & 0.40 & 0.318 (medium) & 0.054 & 0.096 & 0.410 \\
Larynx-cancer stage $\times$ diagnosis year (oncology) & 4$\times$9 & 90 & 0.42 & 0.303 (medium) & 0.042 & 0.090 & 0.369 \\
Cereal maker $\times$ shelf placement (nutrition) & 6$\times$3 & 65 & 0.10 & 0.350 (medium) & 0.210 & 0.233 & 0.376 \\
Partner status $\times$ conformity (social psychology) & 2$\times$3 & 45 & 0.08 & 0.333 (medium) & 0.255 & 0.260 & 0.368 \\
Gears $\times$ carburetors (engineering) & 3$\times$6 & 32 & 0.09 & 0.508 (large) & 0.311 & -- & 0.546 \\
Kidney-infection recurrence $\times$ polycystic disease (nephrology) & 2$\times$2 & 38 & 1.00 & 0.163 (small) & 0.000 & 0.161 & 0.317 \\
\hline
\end{tabularx}
\end{table}

\textbf{When the table is large, the benchmark is too stingy.} The naive V is "small" or even "negligible," a value a reader is taught to ignore, yet it clears its now-tiny noise floor and the test rejects at p $\approx$ 0. The British-doctors smoking cohort is the sharpest case: with 181,467 records, age and smoking status give V = 0.099, which Cohen's table calls "negligible," on an association no one would call negligible. At this sample size the three estimators agree to the third decimal, because the bias that dominates the small tables is gone; what is left is a real effect the benchmark discards.

\begin{table}[htbp]\centering\small
\begin{tabularx}{\textwidth}{>{\raggedright\arraybackslash}X>{\raggedright\arraybackslash}X>{\raggedright\arraybackslash}X>{\raggedright\arraybackslash}X>{\raggedright\arraybackslash}X>{\raggedright\arraybackslash}X>{\raggedright\arraybackslash}X>{\raggedright\arraybackslash}X}
\hline
Table (field) & shape & N & $\chi$\textsuperscript{2} p & naive V (Cohen) & Bergsma & jackknife & floor V\_0.95 \\
\hline
Age $\times$ smoking status (epidemiology) & 5$\times$2 & 181,467 & <10\textsuperscript{-}\textsuperscript{3}\textsuperscript{0}\textsuperscript{0} & 0.099 (negligible) & 0.098 & 0.098 & 0.007 \\
Diamond cut $\times$ clarity (retail) & 5$\times$8 & 53,940 & <10\textsuperscript{-}\textsuperscript{3}\textsuperscript{0}\textsuperscript{0} & 0.143 (small) & 0.142 & 0.142 & 0.014 \\
Wheeze $\times$ age, coal miners (occupational medicine) & 2$\times$9 & 18,282 & <10\textsuperscript{-}\textsuperscript{3}\textsuperscript{0}\textsuperscript{0} & 0.289 (small) & 0.288 & 0.288 & 0.029 \\
Sex $\times$ age group, suicides (public health) & 2$\times$5 & 53,182 & <10\textsuperscript{-}\textsuperscript{3}\textsuperscript{0}\textsuperscript{0} & 0.177 (small) & 0.177 & 0.177 & 0.013 \\
\hline
\end{tabularx}
\end{table}

Two readings, and one rule. \textbf{The point-estimate choice is not cosmetic} where the bias is large: on the 2$\times$2 kidney-recurrence table Bergsma truncates to exactly 0 while the jackknife on the V\textsuperscript{2} scale returns 0.161: one table, two defensible estimators, two answers, the non-U-estimability theorem (Section 3) on real data. And \textbf{the benchmark itself gets the verdict wrong in both regimes}, because the Cohen thresholds are fixed numbers while the honest cutoff, the noise floor, is a function of the table and its sample size (Section 7). Reporting V against a fixed benchmark over-labels the small tables and under-labels the large ones; reporting it against its own floor does neither. Every dataset above is distinct from those used in the companion browser tools. Source: \texttt{rerun/build\_realworld\_examples.py}, \texttt{rerun/m0f\_realworld\_examples.json}, from the Rdatasets public collection.

\section{The interval is a different problem, and it has a better answer}

The interval this paper reports is the guaranteed, unconditional fallback; for an interval that is tighter still, a companion paper (Dwyer 2026c) conditions on the observed margins and pays a change of estimand; that conditional interval runs a third to a half the projected width and, measured directly across 4x4 to 6x6, is near-nominal for moderate-to-large effects and covers the exact-independence null near-nominally (undercovering only small nonzero effects). The claim here is narrower and prior to that one: the interval and the point estimate are different problems, and the square root that ruins the point estimate near the null leaves the interval untouched. The projected interval below is the guaranteed member of the routing rule, and the section that follows separates the two senses of "exact" that make the case.

\subsection{A guarantee is not the same as a tight interval}

Two properties get run together under the word "exact," and separating them is what this section is for. One is the coverage guarantee: does the interval cover at least 1 - $\alpha$? The other is tightness: is it as short as an interval with that guarantee can be? They are independent, and the projected interval below buys the first at a cost in the second.

\begin{table}[htbp]\centering\small
\begin{tabularx}{\textwidth}{>{\raggedright\arraybackslash}X>{\raggedright\arraybackslash}X}
\hline
sense & attainable for the projected interval? \\
\hline
coverage \textbf{= 0.95} for every $\phi$\textsuperscript{2} & \textbf{No.} Discreteness makes attainable coverage jump; no procedure hits 0.95 on the nose. \\
coverage \textbf{$\ge$ 0.95} for every $\phi$\textsuperscript{2} & \textbf{Only as strongly as the set S guarantees it.} The projection covers whatever S covers; here S is the asymptotic likelihood-ratio region, so the guarantee is asymptotic and conservative, not finite-sample exact. \\
coverage \textasciitilde{} 0.95 on average & yes, but it is not a guarantee; this is what the noncentral inversion offers \\
\hline
\end{tabularx}
\end{table}

The projection step is a tautology: p in S implies g(p) in g(S), so the projected interval covers a value of $\phi$\textsuperscript{2} whenever S covers the corresponding p. Whatever coverage S has, the interval has, and no more. An exact finite-sample S, like the Clopper-Pearson binomial region built by inverting exact binomial tails, would make the projected interval exact; the multinomial G\textsuperscript{2} region we use is calibrated to a chi-square and is only asymptotic, so on sparse or near-boundary tables it, and hence the interval, can fall short of the nominal level. What we claim below is therefore the weaker and honest statement: conservative, asymptotically valid coverage, verified in the regimes tested (Section 6.3). The distinction that matters is guarantee against tightness; the Clopper-Pearson binomial interval over-covers and nobody calls it inexact, because the honest word refers to the guarantee, not to the width.

\subsection{A standard construction, borrowed: do not invert a test for $\phi$\textsuperscript{2}}

Inverting a test for $\phi$\textsuperscript{2} requires a supremum over an (RC-1)-dimensional nuisance manifold, because infinitely many p share a $\phi$\textsuperscript{2}. \textbf{That is why the problem has resisted.} The standard evasion is to project, and it is standard:

\textbf{If P(p in S(O)) $\ge$ 1 - $\alpha$ for every p, then for ANY function g, P(g(p) in g(S(O))) $\ge$ 1 - $\alpha$ for every p.}

Because p in S implies g(p) in g(S). \textbf{There is nothing else to prove.}

\textbf{This construction is not ours and we do not claim it.} It is the Scheffe device (Scheffe 1953) applied to a nonlinear functional; it is what Stark's strict bounds do in inverse problems (Stark 1992); it is the projection step of the modern literature on inference for functionals of partially identified parameters (Kaido, Molinari and Stoye 2019); and it is the marginalization of a confidence procedure, a standard graduate-level lemma. Guaranteed intervals for a related functional, mutual information, have been built before, but by a different mechanism, a variational-distance concentration bound (Stefani et al. 2013), rather than by the Scheffe projection of a confidence set, so the projected-interval instantiation here is less anticipated than an earlier draft conceded. \textbf{What we contribute is the instantiation for $\phi$\textsuperscript{2} and Cramer's V, the lower endpoint that is zero in closed form exactly at non-rejection, and the benchmark of Section 6.3 against every interval that current software ships.}

Take S to be the likelihood-ratio confidence region for the multinomial,

\[
S(O) = { p : G^{2}(O; p) \le \chi ^{2}_{(RC-1)}(1 - \alpha ) },
\]

and report [ min\_\{p in S\} $\phi$\textsuperscript{2}(p), max\_\{p in S\} $\phi$\textsuperscript{2}(p) ]. \textbf{No noncentral model. No transform. No calibration.} And it delivers $\phi$\textsuperscript{2}, V\textsuperscript{2}, V, w, T and C simultaneously, because they are all just different g. The price is the Scheffe penalty: the region S is calibrated for the whole (RC-1)-dimensional parameter, so the interval inherits a slack that grows with RC and that no calibration can remove without destroying the guarantee. We state that cost in Section 6.3 rather than bury it.

\textbf{The lower endpoint has a closed form in one case only: it equals zero exactly when the likelihood-ratio test of independence does not reject.} When the lower endpoint is positive it carries no closed form and is solved numerically, like the upper endpoint. The zero case is the noise floor arrived at from the other direction:

\[
min_{p in S} \phi ^{2}(p) = 0 ,\quad \Leftrightarrow  ,\quad G^{2}_{independence}(O) \le \chi ^{2}_{(RC-1)}(1-\alpha ).
\]

\textbf{The interval's lower endpoint is zero exactly when a test of independence fails to reject.} The test, the floor and the interval cannot disagree.

One caveat belongs with the guarantee. The upper endpoint is a maximum of a nonlinear functional over S, and we compute it by numerical optimization with random restarts and no certificate of global optimality. A missed global maximum returns an upper endpoint that is too low, which makes the interval too narrow and could erode its coverage, so the guarantee is contingent on the optimization being accurate and not only on the coverage of S.

\subsection{It works, and here is the bill}

\begin{figure}[htbp]\centering
\includegraphics[width=\linewidth]{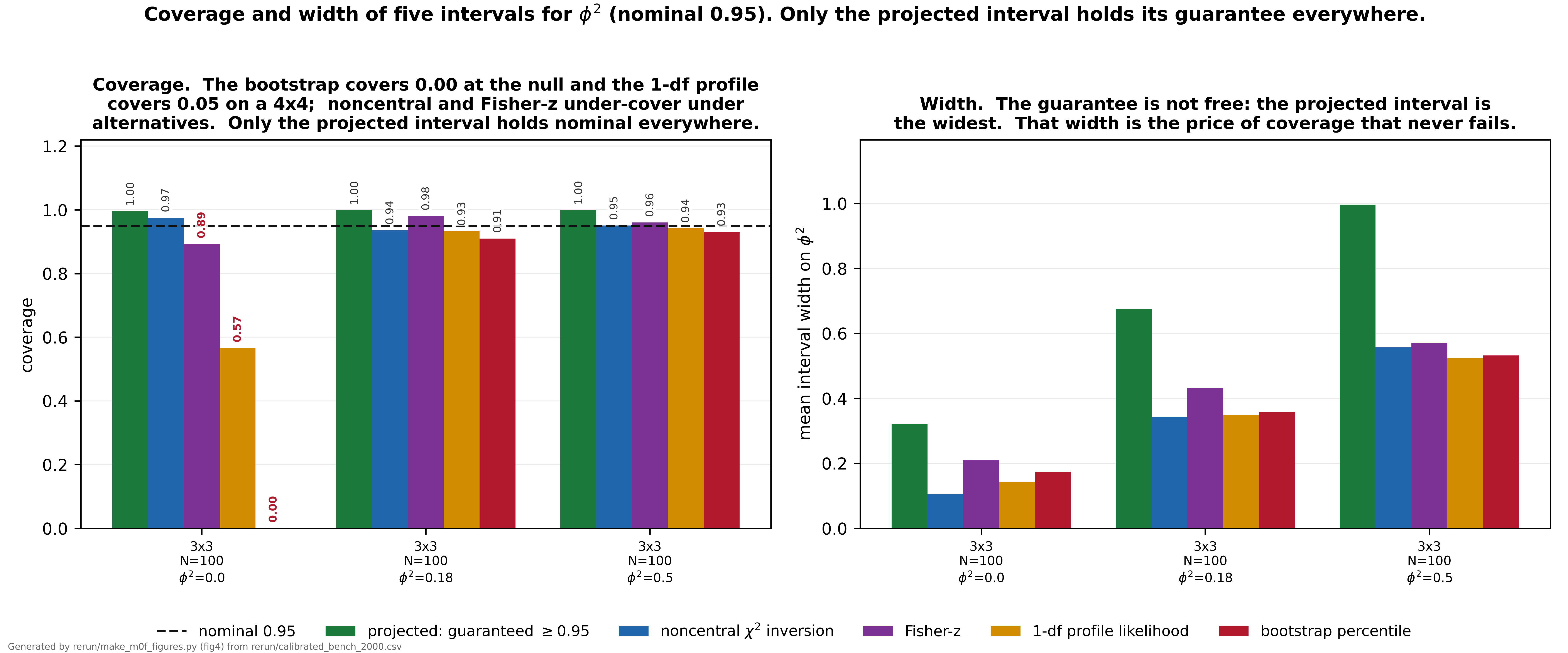}
\caption{\textbf{Figure 5.} Coverage and mean width of the projected interval versus the noncentral chi-square inversion for $\phi$\textsuperscript{2}, over a $\phi$\textsuperscript{2} sweep (3$\times$3, N = 100, 2,000 replications). The projection stays at or above nominal; the noncentral inversion undercovers at moderate effects (0.936 at $\phi$\textsuperscript{2} = 0.18), at two to three times narrower width. Generated by rerun/make\_m0f\_figures.py.}
\end{figure}

\begin{table}[htbp]\centering\small
\begin{tabular}{lccccc}
\hline
design & true $\phi$\textsuperscript{2} & coverage: \textbf{projected} & coverage: noncentral & width: projected & width: noncentral \\
\hline
3$\times$3, N=100 & 0.00 & \textbf{0.997} & 0.975 & 0.321 & 0.106 \\
3$\times$3, N=100 & 0.18 & \textbf{0.999} & 0.936 & 0.676 & 0.342 \\
\hline
\end{tabular}
\end{table}

(2,000 replications per design, from \texttt{rerun/calibrated\_bench\_2000.csv}.)

\textbf{The projected interval holds its coverage; the noncentral inversion does not.} The noncentral inversion over-covers at the null (0.975) and, at $\phi$\textsuperscript{2} = 0.18, covers only 0.936, below the nominal 0.95, so it carries no guarantee even here, while the projected interval stays at or above 0.997. The width ratio is roughly three times at the null and roughly two times at $\phi$\textsuperscript{2} = 0.18, so the guarantee costs roughly two to three times the width. That is the Clopper-Pearson bargain and we state it rather than bury it.

\subsection{The square root breaks the point estimate and leaves the interval untouched}

\begin{table}[htbp]\centering\small
\begin{tabularx}{\textwidth}{>{\raggedright\arraybackslash}X>{\raggedright\arraybackslash}X>{\raggedright\arraybackslash}X}
\hline
 & point estimate & interval \\
\hline
does the square root break it? & \textbf{Yes, at the null.} (+0.097) & \textbf{No. Not anywhere.} \\
why & expectation does not commute with a nonlinear map; Jensen is irreducible & we map a \textbf{set}, not a distribution; a monotone map of a set is exact \\
so report & V\textsuperscript{2}, jackknifed, negatives permitted & the projected interval, \textbf{on whatever scale the reader wants} \\
\hline
\end{tabularx}
\end{table}

\textbf{V may be shown, but only as an interval, never as a bare number.} That is a reporting rule, it follows from a theorem, and it is enforceable in code.

\section{The routing rule}

\begin{figure}[htbp]\centering
\includegraphics[width=\linewidth]{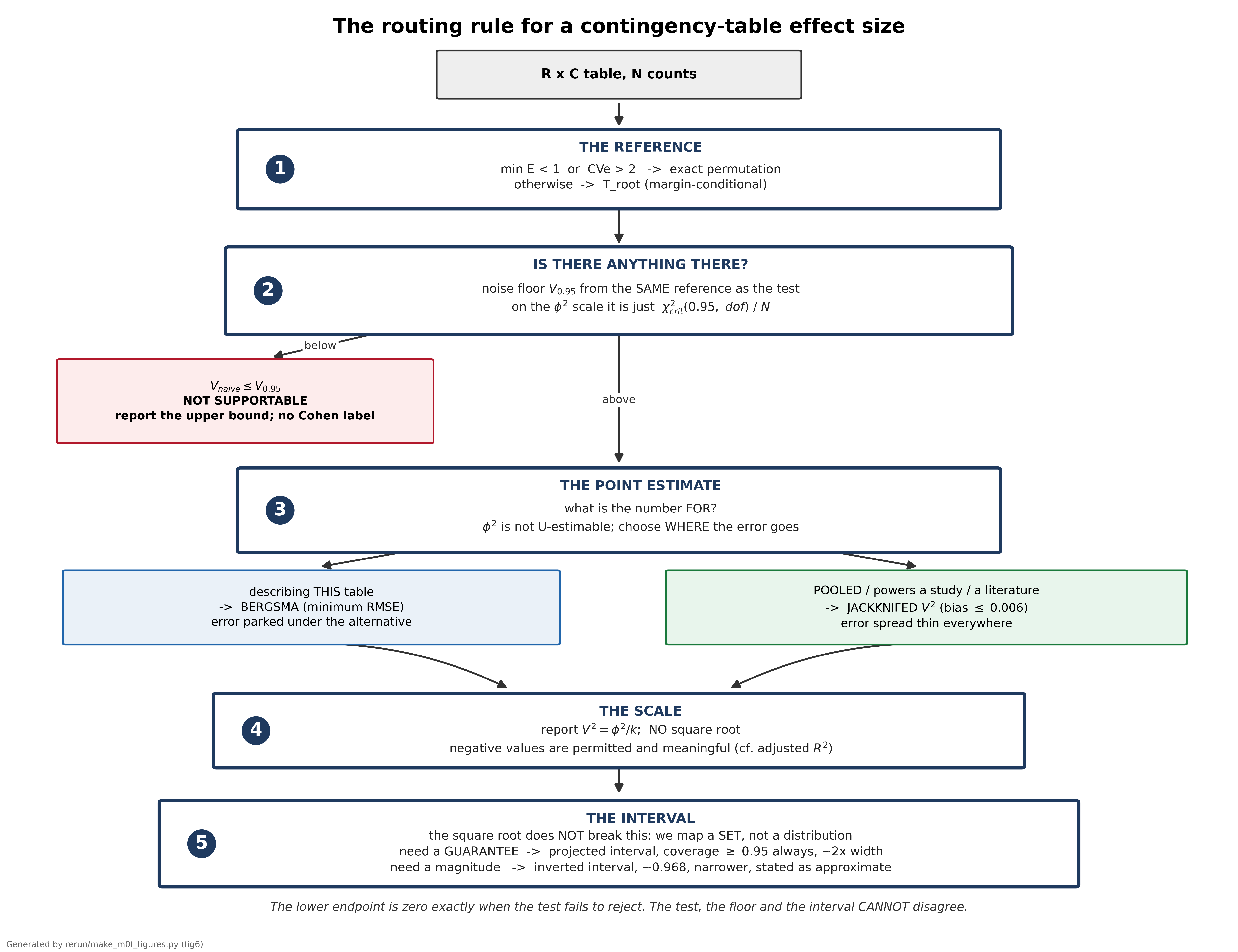}
\caption{\textbf{Figure 6.} The routing rule. A decision flow selecting the reference (by sparsity and margin heterogeneity), the verdict (against the noise floor), the point estimate (by purpose), the scale (always V\textsuperscript{2}), and the interval (by whether a guarantee is required). Generated by rerun/make\_m0f\_figures.py.}
\end{figure}

The M-series already routes the test by sparsity and margin heterogeneity. \textbf{Nothing has routed the effect size, which is exactly how the field arrived here.}

1. \textbf{The reference.} min E < 1 or CVe > 2 $\rightarrow$ exact permutation. Otherwise T\_root, margin-conditional.

2. \textbf{The verdict.} Compare to the noise floor V\_0.95 from the \textbf{same} reference. Below it: not supportable, report the upper bound, apply no Cohen label, stop.

3. \textbf{The point estimate.} The jackknifed V\textsuperscript{2} is the default: it is nearly unbiased everywhere (at most 0.008 at the 180 grid design points, rising to 0.013 as a supremum over the effect axis within the worst design), so it is the right choice whenever the number will be pooled, will power a study, or will enter a literature. Bergsma is the exception, not the rule, earning its place only in the narrow corner of a single table (m = 1) with few degrees of freedom (dof $\le$ 4), where its lower null RMSE wins (derivation D16.4). It is not the minimum-RMSE choice in general: on the Section 5 frontier the shrunk jackknife reaches 0.034 against Bergsma's 0.044.

4. \textbf{The scale.} Always V\textsuperscript{2}. No square root. Negatives permitted and meaningful.

5. \textbf{The interval.} Need a guarantee $\rightarrow$ projected (conservative, asymptotically valid; \textasciitilde{}2x to 3x width). Need a tight interval that is near-nominal rather than merely approximate $\rightarrow$ the exact conditional interval of the companion (Dwyer 2026c), which conditions on the observed margins rather than resting on the noncentral inversion. Need only a rough magnitude and no guarantee $\rightarrow$ the noncentral inversion, which is narrower but undercovers (0.936 at $\phi$\textsuperscript{2} = 0.18) and must be declared approximate.

\subsection{Worked examples}

\begin{table}[htbp]\centering\small
\begin{tabular}{lccccc}
\hline
table & reference & V naive & noise floor V\_0.95 & verdict & V\textsuperscript{2} reported \\
\hline
dense 4$\times$4, N = 300, real effect & T\_root (min E = 13.0) & 0.357 & 0.137 & above & \textbf{+0.118} \\
\textbf{null 5$\times$10, N = 250} & T\_root (min E = 4.1) & \textbf{0.210} & \textbf{0.226} & \textbf{BELOW $\rightarrow$ not supportable} & -- \\
\textbf{sparse 4$\times$4} & \textbf{exact} (min E = 0.89) & \textbf{0.493} & \textbf{0.555} & \textbf{BELOW $\rightarrow$ not supportable} & -- \\
\hline
\end{tabular}
\end{table}

The sparse 4$\times$4 reports \textbf{V = 0.493, which every package labels "large."} The router sends it to the exact reference, finds a floor of 0.555, and refuses it.

\section{A browser tool}

The routing rule is not a diagram to admire; it is a report to produce, and it runs in a browser with no installation and no dependencies. \texttt{honest\_point\_estimate.html} takes a pasted table and returns the whole routed report (the reference chosen by sparsity, the noise floor, the supportability verdict, the point estimate routed by purpose on the V\textsuperscript{2} scale with negatives permitted, and the guaranteed interval), with the reasoning for every decision shown rather than hidden. Figure 7 places that report beside what general-purpose software prints for the same counts.

\begin{figure}[htbp]\centering
\includegraphics[width=\linewidth]{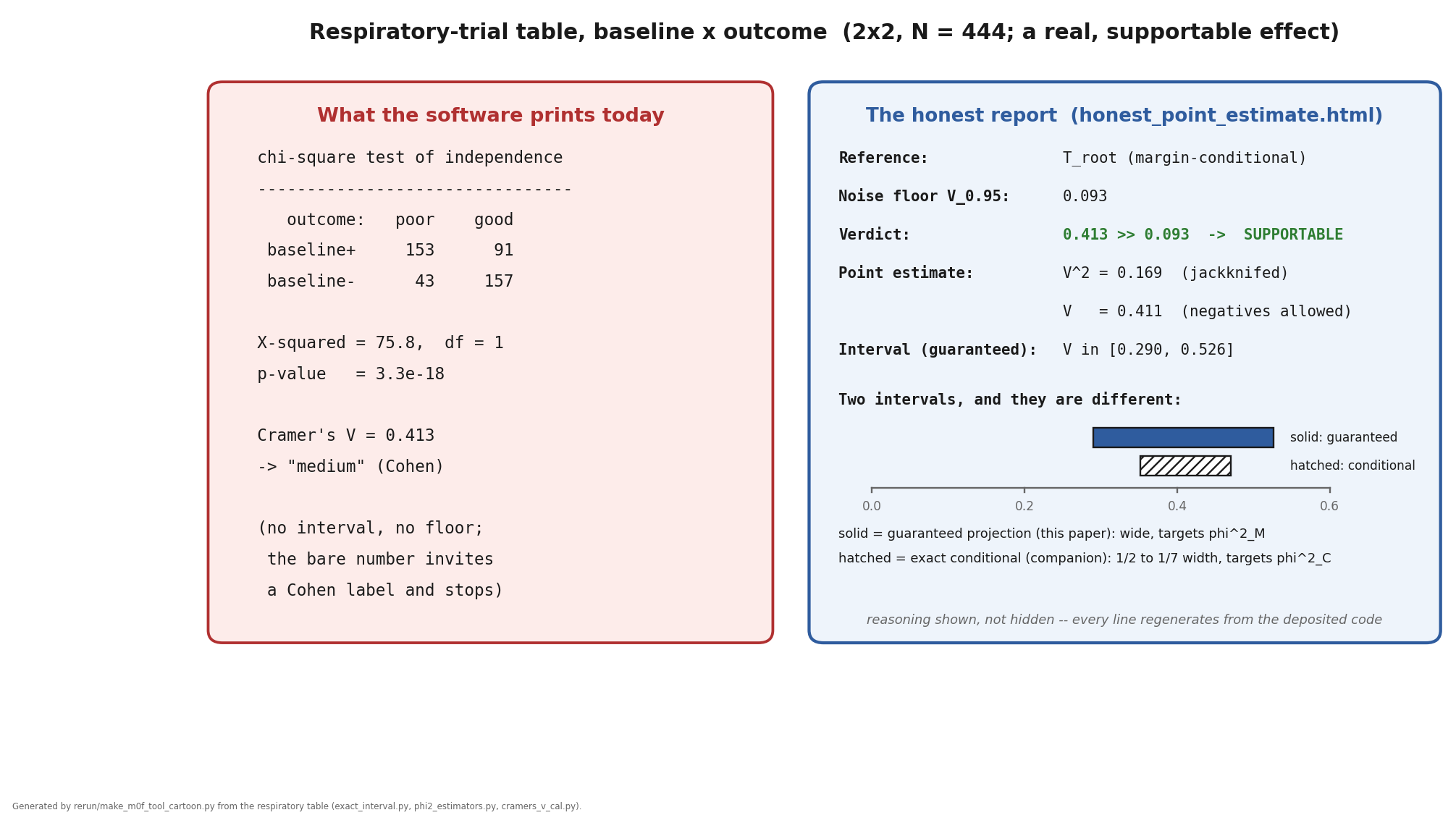}
\caption{\textbf{Figure 7.} The honest report the browser tool produces (right) beside the bare Cramér's V that software prints today (left), for a respiratory-trial table (baseline $\times$ outcome, 2$\times$2, N = 444, a real and supportable effect). Today's output is a single number, V = 0.413, read against a Cohen label; the honest report adds the reference, the noise floor (0.093, cleared), the supportability verdict, the routed point estimate on the V\textsuperscript{2} scale, and the guaranteed projected interval, and it names the two intervals a reader may choose between: the wide guaranteed projection of this paper (targeting the population effect $\phi$\textsuperscript{2}\_M) and the tighter exact conditional interval of the companion (targeting the observed-margin effect $\phi$\textsuperscript{2}\_C). Generated by rerun/make\_m0f\_tool\_cartoon.py.}
\end{figure}

The two intervals are not interchangeable, and the tool says which is which: the projection is guaranteed and wide, the exact conditional is tighter and conditions on the observed margins, so the reasoning behind each choice is on the page rather than in a footnote.

\section{Limitations}

\textbf{We have not removed the bias, because it cannot be removed} (Section 3). The jackknife's residual reaches 0.013 in the worst design we tested. The projected interval costs roughly two to three times the width of the conventional one, and its coverage guarantee is asymptotic and conservative rather than finite-sample exact, because the confidence set it projects is the asymptotic likelihood-ratio region. For small tables its endpoints are found numerically rather than in closed form; the only closed form is the criterion for the lower endpoint being exactly zero (non-rejection), and a positive lower endpoint, like the upper endpoint, is solved by an optimization with random restarts that carries no global-optimality certificate, so a missed maximum would narrow the interval and could erode its coverage. The routing thresholds (min E < 1, CVe > 2) are inherited from the companion test paper and are conventions, not theorems. \textbf{And the choice between Bergsma and the jackknife is a genuine trade that we route rather than resolve, because the theorem says it cannot be resolved.}

\section{Conclusion}

The received practice reports a bare V against a Cohen label. \textbf{We now know three things about that number: it cannot be unbiased under fixed-N multinomial sampling; its square root destroys any correction near the null, which is where most reported effects live; and the interval, which could have carried a stated, conservative guarantee all along, is not being computed.}

The remedy is not a better formula. It is a routed report: the right reference, the floor beside the estimate, the estimate chosen by what it is for, on the bounded affine scale V\textsuperscript{2}, with an interval whose guarantee is stated. \textbf{None of the components is difficult. What was missing was the recognition that they are five different decisions and not one.}

\section*{Declarations}

\textbf{Ethics approval and consent to participate.} Not applicable. This is a methodological and simulation study; it used only simulated data and public example tables, and involved no human participants.

\textbf{Consent for publication.} Not applicable. No individual-person data are reported.

\textbf{Clinical trial number.} Not applicable.

\textbf{Availability of data and materials.} All data are simulated; example tables are from public datasets. The reproducibility package (the effect-size and estimator code, the projected-interval construction, the meta-analysis pooling simulation, locked results, provenance manifest, figures, and the in-browser demonstrator) is openly archived on Zenodo; the concept DOI is 10.5281/zenodo.21783660. Every reported figure and number regenerates deterministically from the deposited \texttt{rerun/} code with fixed seeds.

\textbf{Competing interests.} The author develops and hosts the open-source software and associated web domains (the trialdesign.com applications) that implement the methods described; no other competing interests are declared.

\textbf{Use of generative AI.} In preparing this manuscript the author used a generative-AI assistant (Claude, Anthropic) for drafting and editing prose, generating figure code, and constructing and formatting tables. All AI-assisted output was reviewed and verified by the author; every reported figure and number regenerates deterministically from the openly deposited code, and the author takes full responsibility for the content of this work.

\textbf{Funding.} This research received no specific grant from any funding agency in the public, commercial, or not-for-profit sectors.

\textbf{Authors' contributions.} W. J. Dwyer is the sole author and is responsible for the conception, analysis, software, and writing of this work.

\textbf{Acknowledgements.} Not applicable.

\textbf{Provenance of results.} The numerical results are produced by deterministic, human-reviewed code with fixed seeds; every reported number regenerates from the deposited scripts and locked outputs.

\section*{References}

\noindent Bartlett, M. S. (1937). Properties of sufficiency and statistical tests. \emph{Proc. R. Soc. Lond. A, 160}, 268-282.\par\smallskip

\noindent Bergsma, W. (2013). A bias-correction for Cramer's V and Tschuprow's T. \emph{Journal of the Korean Statistical Society, 42}(3), 323-328.\par\smallskip

\noindent Clopper, C. J., \& Pearson, E. S. (1934). The use of confidence or fiducial limits illustrated in the case of the binomial. \emph{Biometrika, 26}, 404-413.\par\smallskip

\noindent Cohen, J. (1988). \emph{Statistical Power Analysis for the Behavioral Sciences} (2nd ed.). Erlbaum.\par\smallskip

\noindent Cramer, H. (1946). \emph{Mathematical Methods of Statistics}. Princeton.\par\smallskip

\noindent Dwyer, W. J. (2026a). Exact conditional distributions of chi-square-family statistics for two-way contingency tables, by cell-separable dynamic programming. Companion manuscript, submitted for publication.\par\smallskip

\noindent Dwyer, W. J. (2026c). Exact conditional confidence intervals for Cramer's V. Companion manuscript, submitted for publication.\par\smallskip

\noindent Dwyer, W. J. (2026d). An Exact Noise Floor for Contingency-Table Effect Sizes: A Per-Table Reporting Gate, and How Often It Would Change a Reported Magnitude. Companion manuscript, submitted for publication.\par\smallskip

\noindent Fay, R. E. (1985). A jackknifed chi-squared test for complex samples. \emph{Journal of the American Statistical Association, 80}(389), 148-157.\par\smallskip

\noindent Girshick, M. A., Mosteller, F., \& Savage, L. J. (1946). Unbiased estimates for certain binomial sampling problems with applications. \emph{Annals of Mathematical Statistics, 17}(1), 13-23.\par\smallskip

\noindent Halmos, P. R. (1946). The theory of unbiased estimation. \emph{Annals of Mathematical Statistics, 17}(1), 34-43. (Cited here only to disclaim it: see Section 3.)\par\smallskip

\noindent Jiao, J., Han, Y., \& Weissman, T. (2017). Maximum likelihood estimation of functionals of discrete distributions. \emph{IEEE Transactions on Information Theory, 63}(10), 6774-6798.\par\smallskip

\noindent Kaido, H., Molinari, F., \& Stoye, J. (2019). Confidence intervals for projections of partially identified parameters. \emph{Econometrica, 87}(4), 1397-1432.\par\smallskip

\noindent Lehmann, E. L., \& Casella, G. (1998). \emph{Theory of Point Estimation} (2nd ed.). Springer. (Section 2.1.)\par\smallskip

\noindent Momozaki, T., Cho, W., Nakagawa, T., \& Tomizawa, S. (2024). Improving the accuracy of estimating indexes in contingency tables using Bayesian estimators. \emph{Journal of Statistical Theory and Practice, 18}. (arXiv:2109.09339.)\par\smallskip

\noindent Paninski, L. (2003). Estimation of entropy and mutual information. \emph{Neural Computation, 15}(6), 1191-1253. (Proposition 8.)\par\smallskip

\noindent Quenouille, M. H. (1949). Approximate tests of correlation in time series. \emph{JRSS B, 11}, 68-84.\par\smallskip

\noindent Scheffe, H. (1953). A method for judging all contrasts in the analysis of variance. \emph{Biometrika, 40}(1-2), 87-104.\par\smallskip

\noindent Stark, P. B. (1992). Inference in infinite-dimensional inverse problems: discretization and duality. \emph{Journal of Geophysical Research, 97}(B10), 14055-14082.\par\smallskip

\noindent Stefani, A. G., Huber, J. B., Jardin, C., \& Sticht, H. (2013). Confidence intervals for the mutual information. arXiv:1301.5942. (Later in \emph{International Journal of Machine Intelligence and Sensory Signal Processing}, 2014, 1(3), 201-214.)\par\smallskip

\noindent Tschuprow, A. (1925). \emph{Grundbegriffe und Grundprobleme der Korrelationstheorie}. Teubner.\par\smallskip

\noindent Tukey, J. W. (1958). Bias and confidence in not quite large samples. \emph{Annals of Mathematical Statistics, 29}, 614.\par\smallskip

\end{document}